 \newlength\smallfigwidth
 \documentclass[aps,jap,twocolumn,floatfix,showpacs,amsmath,amssymb,superscriptaddress]{revtex4-1}

\usepackage[toc,page]{appendix}
\usepackage{graphicx}
\def\ba{\begin{eqnarray}}
\def\ea{\end{eqnarray}}
\def\be{\begin{equation}}
\def\ee{\end{equation}}

\begin{document}



\title{Berry phases and zero-modes in toroidal topological insulator}

\author{J. M. Fonseca}
\email{jakson.fonseca@ufv.br}
\affiliation{ Departamento de F\'isica, Universidade Federal de Vi\c cosa, 36570-900,
Vi\c cosa, Brazil}

\author{V. L. Carvalho-Santos}
\email{vagson.santos@ufv.br}
\affiliation{Instituto Federal de Educa\c c\~ao, Ci\^encia e Tecnologia Baiano, 
48970-000, Senhor do Bonfim, Brazil}
\affiliation{Departamento de F\'isica, Universidad de Santiago de Chile and CEDENNA, 
Santiago, Chile}

\author{W. A. Moura-Melo}
\email{winder@ufv.br\\ URL: https://sites.google.com/site/wamouramelo/home}
\author{A. R. Pereira}
\email{apereira@ufv.br\\ URL: https://sites.google.com/site/quantumafra/}
\affiliation{ Departamento de F\'isica, Universidade Federal de Vi\c cosa, 36570-900,
Vi\c cosa, Brazil}

\begin{abstract}

An effective Hamiltonian describing the surface states of a toroidal
topological insulator is obtained, and it is shown to support both bound-states and charged zero-modes. 
Actually, the spin connection induced by the toroidal curvature can be viewed as an position-dependent 
effective vector potential, which ultimately yields the zero-modes whose wave-functions harmonically oscillate around the toroidal surface. In addition, two distinct Berry phases are predicted to take place 
by the virtue of the toroidal topology.

%

\end{abstract}

%
%
%

\maketitle

\section{Introduction and motivation}

Topological quantum states of matter do not break any local symmetry, so that they cannot be described by Landau theory of phase transitions. Instead, they demand topological quantum numbers associated with the bulk wavefunctions. Examples include the integer quantum Hall effect, topological insulators \cite{TI review}, topological superconductors and topological superfluids \cite{shen}. On the other hand, geometrical features of the real and momentum spaces are very important to describe the physical properties of several condensed matter systems, like the nematic order in liquid crystals \cite{Napoli-PRL-2012}, graphene \cite{jakson-graphene, Cortijo-NPB-2007} and magnetic systems \cite{C-Santos-JPA-2015,Van-NJP-2009,Vilas-Boas-PLA,C-Santos-JAP,C-Santos-PLA}.\\

Due to their potentialities for application in spintronics and quantum computation, there has been an increasing interest in the study of topological insulators (TI's) in the last years. 
TI's exhibit an insulating bulk while their borders support conducting states protected by time-reversal symmetry (TRS) \cite{TI review}. A TI is characterized by a topological number associated with the Bloch  wavefunctions describing the bulk band structure which can assume two values, $0\,{\rm or}\, 1$, accounting for an ordinary or a topological insulator respectively. In fact, a TI has an odd number of Dirac cones connecting the electronic energy bands on the TI surface and may be described by a two-dimensional Dirac equation \cite{Fan-Zhang, silvestrov}. These cones lie over all orientations of flat surfaces, making them gapless.\\

Actually, curvature has been shown to affect the physical properties of topological insulators. For instance, on a spherical TI, it has been reported that the surface carriers remain gapless while the bound-states experience a gap proportional to the curvature, $1/R$, where $R$ is the sphere radius \cite{spherical TI, spin-connection-TI,Lee}. Furthermore, the effect of the spin connection in surface carriers can be interpreted as fictious magnetic monopole and there are two types of Berry phase for the surface electrons \cite{spherical TI}. 
It is noteworthy to mention that the case of a spherical TI has been considered by \cite{spherical TI}, while a more general situation, describing any curved topological insulator was presented in Ref.
\cite{takane}. Here, we study the toroidal geometry, whose variable curvature leads to an induced spin connection which may be faced as a position-dependent effective vector potential. Furthermore, such a feature yields the zero-modes and two distinct Berry phases to surface carriers.
In addition, a conical TI presents a electric polarization associated to its aperture angle \cite{jakson-cone}, so that narrower or wider cones appear geometrically polarized in opposite directions. Other studies include cylinder \cite{cilindrical-IT} and the M\"obius strip \cite{mobius}, the simplest non-orientable Riemann manifold, where the usual topological invariant is not well-defined. Moreover, the classical limit has been used to study the scattering of electrons in curved TI's with non-trivial metrics \cite{espalhamento-curvo-IT}.\\

Despite the increasing interest in the study of curved TI's, our knowledge is still very superficial with several questions demanding answers. For instance, one knows that the Dirac cones are gapless along all orientations of a flat TI surface\cite{surface-states}. Whether this gapless feature is kept whenever an arbitrary curved TI is considered comes to be an open question, even though the sphere has provided a partial and affirmative answer, as pointed out above. At this point, the toroidal geometry appears to be a very suitable manifold to be studied, once it encompasses a variable curvature interpolating between negative (pseudosphere or hyperbolic plane) and positive (sphere) curvatures. In turn, such a geometry has recently received a considerable attention in the literature. As examples, we may quote toroidal tight traps for Bose-Einstein condensates \cite{Bose-Einstein}, carbon nanotubes providing quasi-zero-dimensional systems whenever the rings are very small \cite{nanotubo1, nanotubo2} and torus-shaped field-effect transistors for tecnological applications \cite{aplic-tecno}. Furthermore, it has been shown that a ferromagnetic nanotori can support a vortex as the magnetization ground state for smaller radius than cylindrical nanorings \cite{C-Santos08, C-Santos10}.\\

Our article is outline as follows: in Section II we present the model and obtain the
surface effective Hamiltonian for a toroidal TI. Section III is devoted to study how the toroidal geometry and topology leads to two distinct Berry phases and the fictitious magnetic monopole induced by the
curvature. Next, in Section IV, we discuss upon the physical spectrum, namely, we show that there is charged zero-modes, a fact brought about by the variable curvature of the toroidal surface. Finally, the conclusions and prospects are presented.
In the appendix we present an alternative derivation of the surface effective Hamiltonian for a toroidal TI (supplementary material of section II).\\
\section{Surface effective Hamiltonian for toroidal topological insulator}

We consider BHZ model describing TI's such as Bi$_2$Se$_3$, Bi$_2$Te$_3$ and Sb$_2$Te$_3$ \cite{zhang-nature, Liu-2010-Hamiltonian}, with a minimal set of physical parameters which encompasses the distinction between an ordinary and a topological insulator. This is done by requiring particle-hole symmetry and isotropic mass term, as below:
\be\label{BHZMINIMAL}
{\cal H}(\vec k)=
\left[\begin{array}{cccc}
M(\vec k)  & A k_z     & 0          & A k_-  \\
A k_z    & -M(\vec k)  & A k_-    & 0        \\
0          & A k_+     & M(\vec k) & -A k_z \\
A k_+    & 0           & -A k_z   & - M(\vec k )\\
\end{array} \right]\,,
\ee
where $k_\pm = k_x\pm ik_y$ and  $M(\vec k) = M - B (k_x^2 + k_y^2 + k_z^2)$; $M/B<0$ concerns for an ordinary while $M/B>0$ accounts for a topological insulator. It is noteworthy that even in this simplest case, the surface states appear whenever there is a topological phase
transition, say, whenever $M/B$ changes its sign at the boundary, for example, 
positive in the bulk and negative in the vacuum. These parameters are computed by fitting the energy 
spectrum of the effective Hamiltonian to that coming from {\it ab initio} calculations\cite{zhang-nature}. In Table \ref{Tabela1} we quote typical values of such parameters for a number of common topological insulators.
\begin{table}[!h]
\begin{tabular}{|c|c|c|c|} \hline
	           & A (eV \AA) & B (eV \AA$^2$) & M (eV \AA)  \\ \hline
 Bi$_2$Se$_3$  & 2.26 $\sim$ 3.33 & 6.86 $\sim$ 44.5  & 0.28  \\ \hline
 Bi$_2$Te$_3$  & 0.30 $\sim$ 2.87 & 2.79 $\sim$ 57.38 & 0.30 \\ \hline
 Sb$_2$Te$_3$  & 0.84 $\sim$ 3.40 & 19.64 $\sim$ 48.51 & 0.22 \\ \hline
\end{tabular}
\caption{Typical values of physical parameters $A$ and $B$ for some TI crystals in the topological insulator phase \cite{TI review}. }\label{Tabela1}
\end{table}

\begin{figure}
\includegraphics[scale=0.7]{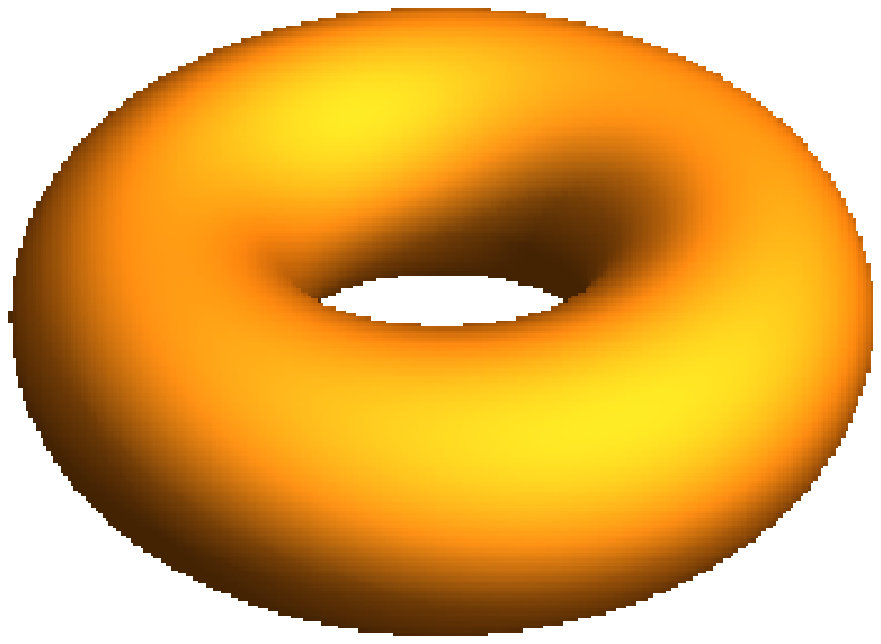}
\includegraphics[scale=0.55]{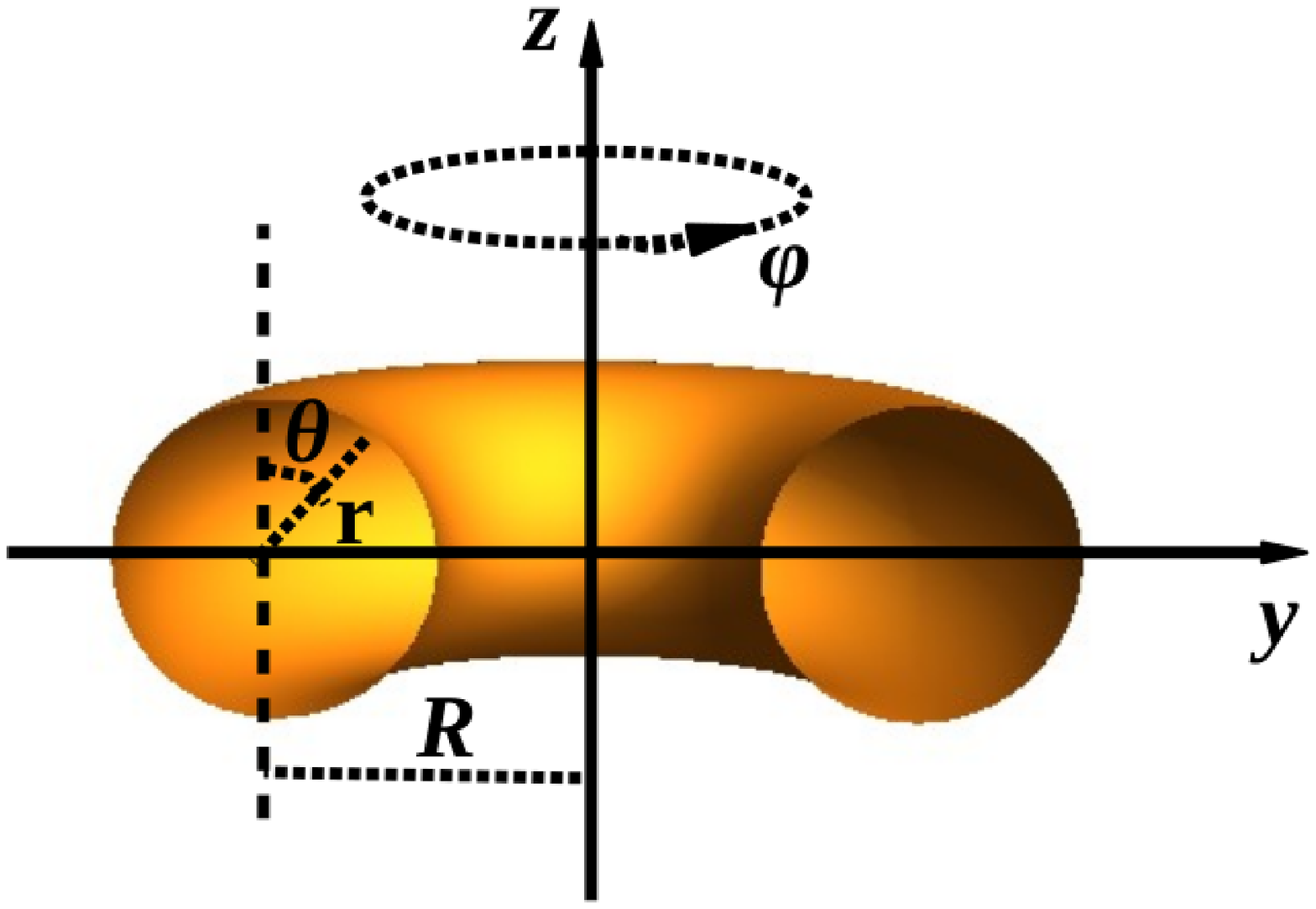}
\caption{[Color online] Above, a ring torus embeded in 3D space is depicted; Below, 
a cut showing the peripolar coordinates, $(\theta, \phi)\,\, \in\,\, [0, 2\pi)$, and the internal and external torus radii, $r$ and $R$.}\label{torus}
\end{figure}
Our interest is to study the above model on the toroidal geometry, which is a smooth
surface with varying curvature. The simplest torus is a surface having genus 
one (a single central hole), and whenever embedded in three dimensional space, it shapes like a donut (see Fig. \ref{torus}). The standard tori are classified in three distinct types concerning the relations between their internal, $r$, and external, $R$, radii. For $R > r$, we have a ring torus (donut shape, as shown in Fig. \ref{torus}), if $R = r$ a horn torus is obtained, while for $R < r$ a spindle torus takes place. Any ordinary torus may be parametrized, for instance, by peripolar coordinates $(\theta,\,\phi)$:

\be
(R-\sqrt{x^2 + y^2})^2 + z^2 =r^2\,,
\ee
which are related to the Cartesian ones by:
\be
\label{parametricequation1}
\left\{\begin{array}{l}
x=(R+r\sin\theta)\cos\phi\,,\\
y=(R+r\sin\theta)\sin\phi\,,\\
z=r\cos\theta\,,
\end{array} \right.
\ee
where $R$ is the rotating (external) and $r$ is the axial (internal) radii, respectively (see Fig \ref{torus}). Thus the momentum operator in Eq. (\ref{BHZMINIMAL}) is written in
toroidal coordinates as: $\vec{p}=p_r\mathbf{\hat{r}}+p_\theta\mathbf{\hat{\theta}}+p_\phi\mathbf{\hat{\phi}}$, 
where $p_r=-i\partial_r$, $p_\theta=-i\partial_\theta/r$ and 
$p_\phi=-i\partial_\phi/(R+r\sin\theta)$.

In what follows, we shall derive a surface Hamiltonian in the $\vec k\cdot \vec p$ approximation. First, note that ${\cal H}(\vec k)$ 
may be split into two parts: ${\cal H}(\vec k)={\cal H}_{\parallel}+{\cal H}_\bot$, 
where ${\cal H}_\bot={\cal H}(\vec k)|_{p_\theta=0,p_\phi=0}$. 
Let us begin, by solving the radial eigenvalue problem:
\be
\label{RadialCompo}
{\cal H}_{\bot}|\psi\rangle=E_{\bot}|\psi\rangle\,.
\ee
Assuming that $|\psi\rangle$ may be written as:
\be
\label{GeneralSolution}
|\psi\rangle=|\psi(r,\theta,\phi)\rangle=e^{\lambda(r-R)}|u(\theta,\phi)\rangle,
\ee
where $\lambda$ accounts for the penetration of the surface wave function into the bulk. Demanding that such a solution satisfies the boundary condition, we must have:
\be
\label{BC radial}
|\psi(r=R)\rangle=0,
\ee
which states that all four components of the wave function vanish on the surface of the torus (at $r=R$). This boundary condition is analogous to that used in the spherical case\cite{spherical TI} 
ensuring that there are no charge carriers outside the TI surface. Boundary condition above demands:
\be
\label{lambda solution}
\lambda_\pm=\frac{-\alpha\beta A\pm\sqrt{A^2+4MB}}{2B}\,,
\ee
\noindent
where $(\alpha, \beta)=\pm1$ specify the spin polarization of the radial eigenvector. Since $\psi$ describes 
localized surface states, $\lambda_\pm$ must be both positive, which demands $\alpha\beta<0$ and $MB>0$, 
since $A/B>0$ (see Table \ref{Tabela1}). In addition, note that the requirement $M/B>0$, coming from 
boundary conditions, must be fulfilled in the topological state \cite{TI review}.\\

Solution of Eq. (\ref{GeneralSolution}) may be written as ($N$ is a normalization constant):
\be
\psi=N[e^{\lambda_+(r-R)}-e^{\lambda_-(r-R)}]\mathbf{u}[\lambda_+]\equiv 
\rho(r)\mathbf{u}[\lambda_+]\,.
\ee
Now, by defining the basis eigenstates of $H_{\bot}$, as $|\psi>=|\mathbf{\pm}\rangle$, where
\be \label{surface-basis}
|+\rangle=\rho(r)|\mathbf{\hat{r}}+\rangle_- \quad
|-\rangle=\rho(r)|\mathbf{\hat{r}}-\rangle_+,
\ee
we obtain:
\be
\label{basis-spinor}
\begin{array}{c}
|\hat{\mathbf{r}}+\rangle_-=\frac{1}{2}
\left[\begin{array}{c}\left[\begin{array}{c}1\\-i\end{array}\right]e^{-i\phi/2}\cos\frac{\theta}{2}\\\\\left[\begin{array}{c}1\\-i\end{array}\right]e^{i\phi/2}\sin\frac{\theta}{2}
\end{array}\right]\\\\
|\hat{\mathbf{r}}-\rangle_+=\frac{1}{{2}}\left[\begin{array}{c}\left[\begin{array}{c}1\\i\end{array}\right]e^{-i\phi/2}\sin\frac{\theta}{2}\\\\-\left[\begin{array}{c}1\\i\end{array}\right]e^{i\phi/2}\cos\frac{\theta}{2}
\end{array}\right].
\end{array}
\ee
The effective surface Hamiltonian acts onto the spinor $|\alpha>$, which in the $\vec k\cdot \vec p$ 
approximation, may be written as $|\alpha\rangle=\alpha_+|+\rangle+\alpha_-|-\rangle$, so that

\be
\left[\begin{array}{c}\langle+|\mathcal{H}_{\parallel}|\alpha\rangle\\\langle-|\mathcal{H}_{\parallel}|\alpha\rangle\end{array}\right]\equiv\mathcal{H}_{s}|\alpha>\quad {\mbox{with}}\quad \mathcal{H}_{s}=A\left[\begin{array}{cc}
0  			& \tilde{{\cal D}}_+    \\
 \tilde{{\cal D}}_-  & 0  
\end{array} \right]\,,
\ee
whi
\begin{eqnarray}
\label{dirac operator torus}
\tilde{{\cal D}}_\pm = \mp  \frac{\partial_\theta}{r} 
\mp \frac{1}{2}\frac{\cos\theta}{R+r\sin\theta}
 +  \frac{i}{R+r\sin\theta}\partial_\phi \,,
\end{eqnarray}
which is the Dirac operator for a free massless fermion on a toroidal surface. It is noteworthy that whenever $R\to 0$ we recover its spherical counterpart \cite{spherical TI}. Instead of using the double-valued basis (\ref{basis-spinor}), we opted by a single-valued one, then we should replace $-i\partial_j$ by $-i\partial_j +1/2$ ($j=\theta, \phi)$.

\section{Berry phases in a toroidal TI}

The two distinct Berry phases appearing on the surface of a toroidal TI are associated to the angles $\theta$ and $\phi$. Indeed, this comes about from the fact that a torus may be generated by the topological product of two circles, say, $S^1_\theta\, \#\, S^1_\phi$ \cite{Nakahara}. To see this in more details, let us note that any solution of
\be\label{eigevaluesurface}
{\cal H}_s\psi(\theta,\,\phi)=E\psi(\theta,\,\phi)\,,
\ee
must satisfy periodic boundary conditions:
\be
\label{BC alpha}
\psi(\theta,\,\phi)=\psi(\theta+2\pi,\,\phi)=\psi(\theta,\,\phi+2\pi)\,.
\ee

Here, $\psi=|\alpha\rangle=\alpha_+|+\rangle+\alpha_-|-\rangle$, and using Eqs. (\ref{surface-basis}-\ref{basis-spinor}), we obtain:
\be
\label{boundary-conditions}
\alpha_\pm(\theta,\phi+2\pi)=\alpha_\pm(\theta+2\pi,\phi)=-\alpha_\pm(\theta,\phi)\,.
\ee
These anti-periodic boundary conditions are equivalent to Berry phases, $\gamma=\pi$, as follows:
\ba
\label{Berry Phase}
\alpha(\theta+2\pi,\phi)=e^{i\gamma_{\theta}}\alpha(\theta,\phi)=e^{i\pi}\alpha(\theta,\phi)\nonumber \\
\alpha(\theta,\phi+2\pi)=e^{i\gamma_{\phi}}\alpha(\theta,\phi)=e^{i\pi}\alpha(\theta,\phi)\,.
\ea
Whenever a charge carrier, whose spin is locked to the surface, completes a closed path along the azimuthal 
or the polar angle, any of such loops cannot be shrunk to a point by means of smooth deformations. 
Physically, its spin is twisted by $2\pi$ in such a way that $\psi$ accumulates a $\pi$-phase. Indeed, TI's 
have a nontrivial intrinsic momentum space Berry phase acquired by an electron when it 
revolves around a Fermi loop for its spin rotates with the momentum $\vec k$ around Fermi 
surface. This is important to describe the electron behavior. For instance, whenever it couples to a 
magnetic field or it is subject to disorder effects, the $\pi$-phase ensures antilocalization, that is, even 
a strong disorder cannot localize surface electronic states, provided that the bulk energy spectrum remains 
gapped.

Alternatively, the Berry phases may be related to the spin connection, $\Gamma_\theta=\frac{1}{2}\frac{\cos\theta}{R+r\sin\theta}$, which is equivalent to a vector potential, $\vec{A}= \Gamma_\theta \hat{\phi}$, generated by an {\it effective magnetic charge} distribution induced by the toroidal curvature, $K=\sin(\theta)/r(R+r\sin(\theta))$. Such a distribution is expected to be a ring of magnetic monopoles, with radius $R$, placed on the $xy$-plane. If we take $R\to 0$, it shrinks to a single Dirac magnetic monopole, located at the sphere center, as obtained in Ref. \cite{spherical TI}. In addition, the curvature appears as a geometrical mechanism to induce artificial gauge fields in TI as much as mechanical tension and topological defects do in the graphene framework\cite{vozmediano physics reports}.

\section{Physical spectrum: bound-states and zero-modes}
The non-trivial curvature of a manifold brings about spin connection on the Dirac equation. Namely, this affects the electronic properties of surface carriers on a toroidal geometry: energy spectrum, spin polarization, local density of states, and so forth. Let us consider the eingenvalue problem (\ref{eigevaluesurface}) with the following ansatz

\be\label{ansatz}
\alpha(\theta,\phi)=e^{im\phi}
\left(\begin{array}{c} \alpha_{m\,+}(\theta)   \\
 \alpha_{m\,-}(\theta) 
\end{array} \right)\,,
\ee
where $m$ accounts for the azimuthal component of the orbital angular momentum, $L_\phi$, taking half-integer values, accordingly to boundary conditions (\ref{boundary-conditions}), say, $m=\pm N/2$ with $N$ an odd integer. Thus, in terms of $\alpha_{m\,\pm}(\theta)$ equation above becomes

\be
\label{alfa+}
\bigg[- \frac{1}{r}\frac{d}{d\theta} 
- \frac{1}{2}\frac{\cos\theta}{R+r\sin\theta}
 - \frac{m}{R+r\sin\theta}\Big]\alpha_{m\,-}=\frac{E}{A}\alpha_{m\,+} \,.
\ee
\be
\label{alfa-}
\bigg[\frac{1}{r}\frac{d}{d\theta} 
+ \frac{1}{2}\frac{\cos\theta}{R+r\sin\theta}
 - \frac{m}{R+r\sin\theta}\Big]\alpha_{m\,+}=\frac{E}{A}\alpha_{m\,-} \,.
\ee
These differential equations may be decoupled at second order, as follow:

\ba
\label{decoupoled equation}
&&\bigg[\frac{1}{(\rho+\sin\theta)}\frac{d}{d\theta}(\rho+\sin\theta)\frac{d}{d\theta}
 - \frac{\sin\theta}{2(\rho+\sin\theta)}\nonumber\\
&&-\frac{1}{(\rho+\sin\theta)^2}\bigg(m+\frac{\sigma}{2}\cos\theta\bigg)^2 +\varepsilon^2
\bigg]\alpha_{m\,\sigma} =0\,,
\ea
where $\sigma=\pm$, $\varepsilon =rE/A$, and $\rho=R/r$. To our best knowledge, this differential equation presents no known analytical solution for $\varepsilon\neq0$ (zero-modes, $\varepsilon=0$, will be discussed below). Efforts have been done to solve it numerically; however, such findings have not been shown useful for a better comprehension of the bound-state spectrum. On the other hand, a qualitative analyses indicates that the energy gap depends on the internal $r$ and external $R$ radii of the torus and it is characterized by two independent quantum numbers, $m$ and $n$, whose discrete energy levels $E(n,m)$ must be symmetrical around $E=0$ (by virtue of the electron-hole symmetric theory). In addition, let us note that as $\rho=R/r \rightarrow 0$ the spherical surface eigenstates along with their spectrum are recovered \cite{spherical TI}, as expected .\\

\indent {\it Zero-modes:} Although surface states on the torus could not be solved analytically for an arbitrary energy, zero-modes solutions may be readily worked out. Actually, by taking $E\equiv0$ in eqs. (\ref{alfa+}) and (\ref{alfa-}), we clearly realize that they are solved by: 
\be\label{zero-modes}
\alpha_0(\theta,\phi)=Ne^{im\phi}
\left(\begin{array}{c} \alpha_{m\,+}(\theta)   \\
 \alpha_{m\,-}(\theta) 
\end{array} \right)\,,
\ee
\be
 \alpha_{m\,+}(\theta) = \frac{1}{\sqrt{\frac{R}{r}+\sin\theta}}
 e^{\frac{2m}{\sqrt{\frac{R^2}{r^2}-1}}\arctan
 \Big(\frac{\frac{R}{r}\tan\frac{\theta}{2}+1}{\sqrt{\frac{R^2}{r^2}-1}}\Big)}\,,
\ee
\be
 \alpha_{m\,-}(\theta) = \sqrt{\frac{R}{r}+\sin\theta}\,
 e^{\frac{2m}{\sqrt{\frac{R^2}{r^2}-1}}\arctan
 \Big(\frac{\frac{R}{r}\tan\frac{\theta}{2}+1}{\sqrt{\frac{R^2}{r^2}-1}}\Big)}\,.
\ee
each of them bearing an integer electronic charge:
\be
Q/e^-=\int d^2r\,j^0=\int d^2r\alpha_0^\dag\alpha_0=1\,.
\ee
The zero-modes 
states appear with wave-functions that harmonically oscillate around the toroidal surface
in the azimuthal and polar directions ($\phi$ and $\theta$ coordinates).
To explain the physical origin of these zero-modes  we recall that the toroidal 
curvature  is equivalent to an {\em effective vector potential}, $\vec{A}=\frac12 \frac{\cos\theta}
{R+r\sin\theta} \hat{\phi}$, as pointed out in the previous section. Now, let us consider a given zero-mode 
displacing a loop along $\phi$, in such a way that it experiences an effective magnetic flux given by 
(recall that $d\vec{l}=dr \hat{r} + r d\theta\,\hat{\theta} + (R+r\sin\theta)d\phi\, \hat{\phi}$):

$$\Phi=\oint \vec{A}\cdot d\vec{l}= \pi \cos\theta .$$

Therefore, a closed path along $\phi$ with constant curvature, one realizes a zero-mode state, 
labeled by a quantum number $ m$. 
The curvature then plays a role analogous to an external field axially applied to a TI cylinder, where it is 
induced a pair of zero-modes depending on the magnetic flux \cite{Zhang-PRL-105-206601-2010}. Furthermore, 
once the torus curvature smoothly varies as $\theta$ goes from $0$ to $2\pi$, it offers a unique framework 
where zero-modes appear due to the geometry of the surface. 

For the sake of completeness, it is worthy to mention that the emergence of zero-energy modes can be established, a priori, by mathematical index theorems, which relates the occurrence of such modes to the geometry and topology of the space on which the Dirac equation is stated \cite{Ansourian77,index}. For instance, a topologically non-trivial background or a defect may bring about a position-dependent mass term into the Dirac equation enabling the spectrum to support isolated zero-modes \cite{jackiw-arxiv-2011,Jackiw84}. Charged zero-modes can be also found in a topological insulator when its surface is coated with a ferromagnetic film supporting an in-plane magnetic vortex. In this case, they have the additional property of being fully polarized modes located near the vortex center \cite{jaksonh2}.

\section{conclusions}

We have shown that the surface of a toroidal topological insulator supports zero-energy modes, while the bound-states have an energy quantization depending on two independent quantum numbers, each one associated with a distinct circle composing the torus by means of a 2-circle topological product, $T^2=S^1 \# S^1$. Actually, the torus variable curvature leads to an additional contribution to the surface states eigenvalue problem, preventing us from obtaining analytical solutions with non-vanishing energy.

On the other hand, we have realized that it is precisely this variable curvature which enables the zero-energy spectrum to appear, as the case of a TI cylinder subject to an applied magnetic field \cite{Zhang-PRL-105-206601-2010}. In our framework, the curvature mimics the effect of a vector potential, yielding non-trivial {\em effective magnetic flux} to the surface electrons whenever they displace along a closed loop along the torus major radius, say, angle $\phi$. Therefore, for each closed path with a fixed curvature, the modes evolve as if they were experiencing a given flux. Therefore, our findings suggest geometrically non-trivial manifolds as useful frameworks for the study of how the curvature (and perhaps other ingredients, like torsion) affects the physical properties of surface states in topological insulators.

In practice, the best fabricated samples are still imperfect in geometry, 
pureness, and so forth. Even for a realistic toroidal TI, where (small) imperfections and non-magnetic 
impurity concentration take place, our main results are not expected to be jeopardized. For instance, the 
zero-modes surface states must be kept, once it relies on more general considerations, say, the torus topology and its 
smoothly variable curvature.

As a final remark, it should be stressed that it is possible to construct a tight-binding model, taking into account nearest-neighbor interaction, as it has been done for the spherical case. By solving such a Hamiltonian, its physical spectrum may be employed even for very small systems, say, nanoscaled torii, presumably with the similar properties as found here.

\vskip .5cm
\centerline{\bf Acknowledgments\\}

The authors thank O.M. Del Cima and D.H.T. Franco for useful discussions and R.J.C. Lopes for
computational help. They are also grateful to CNPq, FAPEMIG and CAPES (Brazilian agencies) 
for financial support.\\

{\it Author Contributions}: All the authors contributed equally to the article.\\

\appendix \section{Dirac equation on a torus}

In order to confirm the results presented in Section II, we shall obtain the Dirac equation for 
a toroidal TI following the general method from Ref. \cite{takane}. There, it is realized that in a curved surface the Dirac equation is modified by the spin connection. On the other hand, a genus-1 torus, Fig. \ref{torus}, is a compact surface whose Gaussian curvature is given by $K=\frac{\sin\theta}{r(R+r\sin\theta)}$, smoothly varying from $-1/r(R-r)$ to $+1/r(R+r)$ along its polar $(\theta)$ angle, so interpolating between the pseudospherical and spherical curvatures whenever $R>r$, and vanishing when $\theta=0$ or $\theta=\pi$. Despite the Dirac equation on a torus can be written using the covariant formalism of quantum field theory in curved space time, here we adopt the general approach of Ref. \cite{takane} which is equivalent to the method presented in the previous sections.
In Ref. \cite{takane} the effective Hamiltonian for the two-component spinor $\alpha(\theta,\,\phi)$ is given by:

\be
\label{surfaceD}
\mathcal{H}_{s}=\left[\begin{array}{cc}
0  			& {\cal D}_+    \\
 {\cal D}_-  & 0  
\end{array} \right]\,,
\ee
where
\begin{eqnarray}
\label{dirac operator D+}
{\cal D}_+ = \sum_{i=1}^{2}\Big[(a_iA&-&b_iB)\Big(\partial_i +
 \frac{1}{2}\partial_i\ln \langle\sqrt{\cal G}\rangle\Big) \nonumber \\ 
& + & \frac{1}{2}[\partial_i(a_iA-b_iB)\Bigg]\,,
\end{eqnarray}

\begin{eqnarray}
\label{dirac operator D-}
{\cal D}_- = \sum_{i=1}^{2}\Big[-(a_iA&-&b_iB)^*\Big(\partial_i +
 \frac{1}{2}\partial_i\ln \langle\sqrt{\cal G}\rangle\Big) \nonumber \\ 
& - & \frac{1}{2}[\partial_i(a_iA-b_iB)^*\Bigg]\,.
\end{eqnarray}
The parameters $a_i$ and $b_i$ are determined by the torus geometry and are given by:
\[
a_1=-\frac{1}{r}\,,\quad a_2 = \frac{i}{R+r\sin\theta}\,,
\]
\[
b_1=-\frac{1}{r^2}\,,\quad b_2 = -\frac{i\sin\theta}{(R+r\sin\theta)^2}\,.
\]
Assuming that the penetration depth $\lambda$ for the surface states is much shorter than $r$, then $\langle r \rangle $ can be approximated by the surface radius and 
$\langle\sqrt{\cal G}\rangle = r(R+r\sin\theta)$ (see Ref. \cite{takane} for more details).
Therefore, Dirac operator ${\cal D_\pm}$ reads:
\begin{eqnarray}
\label{dirac operator renormalizable}
{\cal D}_\pm = \Big(A+\frac{B}{r}\Big)\Bigg(&\mp & \frac{\partial_\theta}{r} 
\mp \frac{1}{2}\frac{\cos\theta}{R+r\sin\theta} \nonumber \\ 
& + & \frac{i(A-B\sin\theta)}{R+r\sin\theta}\partial_\phi\Bigg)\,,
\end{eqnarray}
where $a_iA-b_iB$ may be faced as an effective velocity in
$x^i=(\theta,\,\phi)$ direction; $B/r$ is a (renormalization) correction to the effective velocity of the surface carriers, while $B\sin\theta$ accounts for the renormalization due to curvature. However, these terms can be ignored at first approximation, in such a way that Dirac operator (\ref{dirac operator renormalizable}) comes to be identical to the surface Hamiltonian (\ref{dirac operator torus}),  derived in Section II.

It should be emphasized that a factor $\sqrt{\cal G} = r(R+r\sin\theta),$ (${\cal G}={\rm det} \{g_{ij}\}$, $g_{ij}$ is the torus metric) must be introduced in the definition of the matrix elements of the Hamiltonian (\ref{surfaceD}) whenever evaluated between surface states, because in a 2D manifold
the natural integral measure is $dS=\sqrt{\cal G}dx^1dx^2=r(R+r\sin\theta)d\theta d\phi$.
The correction term $\frac{1}{2}\frac{\cos\theta}{R+r\sin\theta}$ in the above linear differential operator corresponds to a spin connection in the Dirac theory on curved surfaces \cite{takane}, describing the effect (coupling) of the curvature on electrons motion.


\end{document}